# Light up that Droid! On the Effectiveness of Static Analysis Features against App Obfuscation for Android Malware Detection

Borja Molina-Coronado, Antonio Ruggia Usue Mori, Alessio Merlo,
Alexander Mendiburu and Jose Miguel-Alonso

———————————— ✦ ————————————

**Abstract**—Malware authors have seen obfuscation as the mean to by-pass malware detectors based on static analysis features. For Android, several studies have confirmed that many anti-malware products are easily evaded with simple program transformations. As opposed to these works, ML detection proposals for Android leveraging static analysis features have also been proposed as obfuscation-resilient. Therefore, it needs to be determined to what extent the use of a specific obfuscation strategy or tool poses a risk for the validity of ML malware detectors for Android based on static analysis features. To shed some light in this regard, in this article we assess the impact of specific obfuscation techniques on common features extracted using static analysis and determine whether the changes are significant enough to undermine the effectiveness of ML malware detectors that rely on these features. The experimental results suggest that obfuscation techniques affect all static analysis features to varying degrees across different tools. However, certain features retain their validity for ML malware detection even in the presence of obfuscation. Based on these findings, we propose a ML malware detector for Android that is robust against obfuscation and outperforms current state-of-the-art detectors.

## 1 INTRODUCTION

W ITH the spread of Android devices, the amount of malware crafted for this OS has also experienced a extraordinary growth [1, 2]. This has led researchers to devise cutting-edge anti-malware solutions based on machine learning (ML) algorithms. When fed with app data, these algorithms are able to find patterns that are characteristic

- *Borja Molina Coronado, Alexander Mendiburu and Jose Miguel-Alonso are with the Dept. of Computer Architecture and Technology, University of the Basque Country UPV/EHU, Donostia, Spain.*
  *E-mail:{borja.molina, alexander.mendiburu, j.miguel}@ehu.es*
- *Usue Mori is with the Dept. of Computer Science and Artificial Intelligence, University of the Basque Country UPV/EHU, Donostia, Spain.*
  *E-mail: usue.mori@ehu.es*
- *Antonio Ruggia is with the Dept. of Informatics, Bioengineering, Robotics and Sytems Engineering, University of Genoa, Genoa, Italy.*
  *E-mail: antonio.ruggia@dibris.unige.it*
- *Alessio Merlo is with the CASD - Centre for Advanced Defense Studies, Rome, Italy.*
  *E-mail: alessio.merlo@ssuos.difesa.it*
- *Corresponding author: Borja Molina-Coronado*

and informative enough to classify apps as either goodware or malware. In this sense, the performance of ML highly depends on the quality and soundness of the data that is used to build the classifier [3, 4]. In the case of Android malware detection, the extraction of this data, in the form of a vector of features that represents the behavior of apps, is performed using either dynamic or static analysis [5, 6].

Dynamic analysis is performed on a controlled environment (sandbox) where the app is executed. During execution, traces that describe the behavior of the app, e.g., network activity, system calls, etc. are logged [5]. On the contrary, static analysis is based on the inspection of the content of the package file (APK) of an app. This includes the compiled code and other resources such as image and database files [7]. Both techniques are valid to extract valuable data from apps. However, dynamic analysis involves a costly process whose success is dependent on the emulation method used and the absence of sandbox evasion artifacts in the code of apps. Instead, static analysis is computationally cheaper, but it can be counteracted by applying app code transformations. Such transformations are commonly known as obfuscation [8].

Obfuscation is a security through obscurity technique that aims to prevent automatic or manual code analysis. It involves the transformation of the code of apps, making it more difficult to understand but without altering its functionality [9]. This characteristic has made obfuscation a double edged sword, used by both, goodware and malware authors. Developers of legitimate software leverage obfuscation to protect their code from being statically analyzed by third parties, e.g., trying to avoid app repackaging or intellectual property abuses [10]. Malware authors have seen obfuscation as a mean to conceal the purpose of their code [11], preventing static analyses from obtaining meaningful information about the behavior of apps.

It may seem common sense that the application of any, or the combination of several, obfuscation techniques will make malware analysis relying on features extracted using static analysis fruitless. However, it is unclear to what extent this aspect is true. Some studies on Windows and Android executables have demonstrated that obfuscation harms detectors that rely on static analysis features. For example, packing[1] prevents obtaining informative features [12, 13], which are essential to train classifiers. Similar conclusions have been drawn for other forms of transformation [14, 15], showing a major weakness in Android malware detectors. However, other studies contradict what has been stated in the aforementioned works, proposing feature extraction techniques via static analysis that enable a successful identification of malware even when apps are obfuscated [16–18].

All of these works appear promising in demonstrating either the flaws or the strengths of static analysis features for malware detection. However, these discrepancies complicate the extraction of sound conclusions regarding the validity of static analysis features for Android malware detection. In addition, many of these works focus solely on the labels predicted by the detectors, without analyzing the effect of the obfuscation on the apps and/or features used to train them [14, 15, 17, 19, 20]. This additional feature-centered information is important to understand and explain why the detectors are working or failing when obfuscation is present, and is crucial for building more robust detectors. Finally, another evident flaw of some of these studies is the lack of details concerning their datasets and the configuration of their experimental setups [16, 18, 21, 22]. Apart from the lack of reproducibility, biases in the datasets may lean the results towards non-generalizable results. Therefore, the conclusions drawn from all these works may have limited applicability beyond the evaluated scenarios, and can be the cause of the contradictions found in the literature.

To the best of our knowledge, this work presents the first comprehensive study about the impact of common obfuscation techniques in the information that is obtained through static analysis to perform malware detection with ML algorithms. The contributions of this paper can be summarized in the following highlights:

- We provide an agnostic [2] evaluation of the strength, validity and detection potential of a complete set of features obtained by means of static analysis of APKs when obfuscation is used.
- We analyze the impact of a variety of obfuscation strategies and tools on static analysis features, providing insights about the use of these features for malware detection in obfuscated scenarios.
- We propose a high-performing ML-based Android malware detector leveraging a set of robust static analysis features. We demonstrate the ability of this detector to identify goodware and malware despite obfuscation, outperforming the state-of-the-art.
- We present a novel dataset with more than 95K obfuscated Android apps, allowing researchers to test the robustness of their malware detection proposals.
- In spirit of open science and to allow reproducibility, we make the code publicly available at gitlab-borja.

The rest of this paper is organized as follows. Section 2 introduces the literature that has previously tackled obfuscation as a problem in malware analysis. Section 3 provided basic information about topics that are required to understand the content of this paper. Section 4 describes the construction of the app dataset and presents the features that are considered in our experiments. Section 5 evaluates the impact of different obfuscation strategies and tools in static analysis features, as well as their validity for malware detection. Section 6 is devoted to assess the robustness of our ML malware detection proposal. Section 7 includes a discussion of the main findings made along this paper. Finally, we conclude this paper in Section 8.

## 2 RELATED WORK

The related work can be divided into two groups: (1) studies that analyze the vulnerabilities of malware detectors when obfuscation is present, and (2) works that propose novel malware detectors which are presumably robust to obfuscation.

### 2.1 Study of the Vulnerabilities of Malware Detectors

The works that evaluate the negative effects of obfuscation on Android malware detectors have mainly been carried out for black box malware detectors, i.e., the system or model is analyzed and evaluated based solely on its input-output behavior, without direct access to or knowledge of its internal workings. The first work of this type [19] studied how obfuscation impacts the detection ability of 10 popular anti-virus programs available in the VirusTotal platform. The work demonstrated that these detectors are vulnerable and loose their reliability in the identification of obfuscated malware. Similarly, in [20], 13 Android anti-virus programs from VirusTotal are assessed using different obfuscation strategies to modify malware. The results showed a meek improvement in detection accuracy concerning the findings of previous works [19] and proved that companies responsible of developing these tools are trying to counteract obfuscation. A more comprehensive analysis for 60 anti-virus tools in VirusTotal has been presented in [14]. Again, the work demonstrated the vulnerabilities of most detectors when facing obfuscated malware. However, this analysis shows that the success on bypassing detection highly depends on the obfuscation tools and strategies considered.

In the mentioned studies, the detectors are commercial products with unknown characteristics. Some other works have focused on assessing the impact of obfuscation in published ML based detectors. In [17], an analysis of the effect of obfuscation in two detectors, one relying on static and the other on dynamic analysis features, is presented. It is shown that the performance of the detector using dynamic analysis features is not altered by obfuscation, contrary to the detector that uses static analysis features. However, authors indicated that this effect can be easily mitigated by including obfuscated samples during the training phase of ML models. In [15], eight state-of-the-art Android malware detectors leveraging static analysis features and ML algorithms are assessed using obfuscated malware samples. The authors demonstrated that obfuscation is a major weakness of these popular solutions, since all of them suffered a drop in their performance. One of the most recent and comprehensive studies is carried out in [12]. This work analyzes the effect of packing in ML malware detectors relying on static analysis for Windows executables. The conclusions drawn from the extensive set of experiments indicate that ML malware detectors for Windows fail to identify the class of transformed samples due to the insufficient informative capacity of static analysis features.

---

1. Packing is a particular form of obfuscation which hides the real code through one or more layers of compression/encryption. At runtime, the unpacking routine restores the original code in memory to be then executed.

2. In this context, we refer agnostic as an analysis carried out without focusing on a specific malware detection proposal.

All these works prove the added difficulty that obfuscation entails for malware detection. However, most of them fail to provide explanations behind accurate or erroneous detections. In this sense, they treat the detectors as black-box tools and do not analyze the effect of different obfuscation strategies and tools on the apps and, specifically, on the features that will be used for training the detectors. This makes it difficult to extract meaningful insights and provides no useful information to build more robust classifiers.

## 2.2 Obfuscation-Resilient Detectors

A second group of proposals focuses on the development of obfuscation-resilient detectors, specifically designed to operate effectively in the presence of obfuscated apps. Two of the most relevant works in this regard are DroidSieve [16] and RevealDroid [18]. The former categorizes static analysis features as obfuscation-sensitive and obfuscation-insensitive based on theoretical aspects. Feature frequency is studied for different datasets with obfuscated and un-obfuscated malware samples to support the idea that most changing features provide better information. In consequence, they proposed a detector that relies on the features of both groups, and offering good performance in terms of malware detection and family identification. The latter work argues against static analysis features such as Permissions, Intents or Strings for robust malware detection. Contrary to the authors of DroidSieve, they suggest that obfuscation-sensitive features do not provide useful information to detect malware. Instead, the authors propose a new set of static analysis features based on a backward analysis of the calls to dynamic code loading and reflection APIs. In this way, the functions invoked at runtime are identified, nullifying the effect of obfuscation, making the proposed detector obfuscation-resilient.

Two allegedly obfuscation-resilient detectors leveraging deep learning algorithms are presented in [21] and [22]. The authors of these works suggest that the capacity of deep learning to embed and extract useful information from the features is enough to tackle obfuscation. The first work relies on strings extracted from the app code. Strings are then transformed into sequences of characters to obtain an embedded representation of the app that is then used for classification. Despite the excellent results reported for malware detection, the ability of the detector to identify obfuscated apps is based on (unproven) statements that are not specifically tested. The latter proposal incorporates obfuscation-sensitive and insensitive features, including permissions, opcodes and meta-data from ApkID [3], a signature-based fingerprinting tool. Similarly to the previous proposal, the obfuscation-resiliency of this work cannot be confirmed based on the results, since the effect of the use of obfuscation in the detector is based on theoretical aspects not specifically covered by the experiments.

The experiments carried out in all these works present some flaws that, in our opinion, put in question their capability. For example, most of them do not describe, or vaguely analyze, the composition of their datasets in terms of the number of obfuscated malware or goodware samples, as well as the tools and strategies considered to obfuscate the samples. Some articles focus their analyses exclusively on obfuscated malware, either for the training or evaluation of the detectors, but what about obfuscated goodware? How do detectors behave in the presence of such apps? The use of different obfuscation tools or strategies for malware and for goodware introduces biases in ML algorithms, since the generated models may associate obfuscation, or the use of a particular obfuscation tool, to a specific class in the data [12]. Additionally, experiments performed with malware and goodware captured from different periods can cause biases in the detectors [23]. Also, most of these studies focused on a small set of features, arguing against other types of features without providing any proof. All these aspects may justify the good published results and cause contradictions concerning other analyses carried out for ML-based detectors [15, 17]. Finally, we also found that most of them do not provide enough details to reproduce their systems and thus, lack of reproducibility.

## 3 BACKGROUND

This section briefly introduces some basic concepts needed to understand the rest of this paper. This includes the structure and content of an Android Application Package (APK) from which static analysis features are extracted, and the types of obfuscation techniques and their effect in the apps.

### 3.1 Android Apps

Android apps are usually developed in Java or Kotlin[4]. When an app has to meet very strict performance constraints, or interact directly with hardware components, Android allows developers to introduce native components written in C and C++ (i.e., *native code*). An Android app is distributed and installed via an APK, a compressed (ZIP) file containing all the resources needed (e.g., code, images) to firstly execute the app. Figure 1 shows the internal structure of an APK file.

Every APK must be signed with the private key of the developer. To validate this signature, the APK contains the public certificate of the developer inside the `META-INF` folder. This mechanism guarantees the integrity of the APK[5]. In a nutshell, before installing an app, Android verifies if the files in the APK match a pre-computed signature and continues with the installation only if the integrity check succeeds.

The `AndroidManifest.xml` defines the structure of an Android app and its meta-data, such as the package name of the app, the required permissions, and the main components (i.e., Activity, Service, Broadcast Receiver, and Content Provider). An Android app can contain one or multiple DEX file(s) (i.e., `classes*.dex`), which include the compiled Java code. Each `.dex` file can reference up to 64k methods [24], such as the Android framework methods, other library methods, and the app-specific methods. For the native components, Android provides an Android Native Development Kit (NDK) [25] that generates native libraries in the form of Linux shared objects. Such objects are stored into the `lib` folder.

---

3. https://github.com/rednaga/APKiD

4. From now on, we will refer to Java code, although the techniques we describe are also valid for apps written in Kotlin.

5. Note that Android does not verifies the validity of the developer's certificate but instead, uses this mechanism to validate the integrity of the content within the APK. Therefore, the developers' certificates can be self-signed.

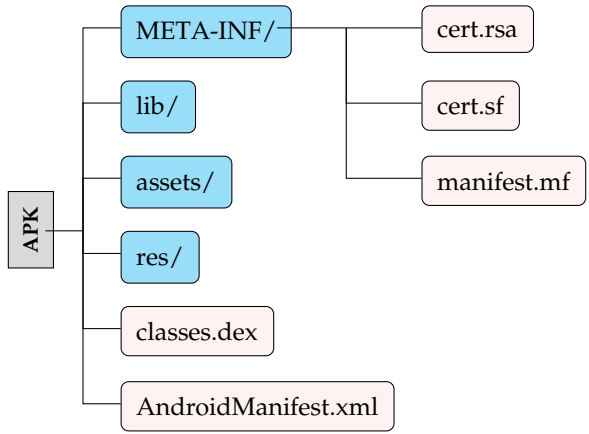

Fig. 1: Structure of an APK file

Finally, the `res` folder contains the compiled resources (e.g., images, and strings), and the `assets` directory includes the raw resources, providing a way to add arbitrary files such as text, HTML, font, and video content into the app.

## 3.2 Obfuscation

Obfuscation is the process of modifying an executable without altering its functionality [26]. It aims to counteract automatic or manual code analysis. In the Android context, many strategies can be applied to modify the code or resources within the APK file: from simple operations that change some metadata to bypass basic checks (e.g., signature-based anti-malware), to techniques that explicitly modify the DEX code or resources of the app [27]. It is worth emphasizing that in Android obfuscation is more common than in other binary code (e.g., x86 executables), because analyzing and repackaging an Android app is straightforward [13]. In the rest of this Section, we present the type of modifications considered in this work.

3.2.0.1 Renaming: A DEX file stores the original string-valued identifiers (names) of fields, methods and classes [28]. Often, these identifiers leak information about code functionalities, lifecycle components and how they interact with each other. For instance, a common practice by programmers is to add "Activity" to each Java class that implements an activity component. The renaming technique replaces these identifiers with meaningless strings, aiming to remove information about the functionality of the app. Consequently, renaming involves modifying the .dex files and the Manifest file (`AndroidManifest.xml`). Note that this technique cannot be applied to methods of the Android lifecycle (e.g., onCreate, onPause) or Android framework components because that would break the execution logic.

3.2.0.2 Code manipulation: These techniques manipulate the DEX code to remove useless operations, hide specific API invocations, and modify the execution flow. The main techniques in this category are:

- *Junk code insertion (JCI)* This technique introduces sequences of useless instructions, such as nop (i.e., *no-operation* instructions that do nothing). Other JCI strategies transform the control-flow graph (CFG) of apps by inserting `goto` instructions or arithmetic branches. For example, a `goto` may be introduced in the code pointing to an useless code sequence ending on another `goto` instruction, which points to the instruction after the first `goto`. The arithmetic branch technique inserts a set of arithmetic computations followed by branch instruction that depends on the result of these computations, crafted in such a way that the branch is never taken [29].

- *Call indirection (CI)* This technique aims to modify the call graph and, therefore, the CFG of the app. It introduces a new intermediate chain of method invocations in the code, adding one or several nodes between a pair of nodes in the original graph. For example, given a method invocation from $m_{or1}$ to $m_{or2}$ in the code, $m_{or1}$ is modified to call to the start of a sequence of $n$ intermediate methods ($m_i$ : $1 <= i <= n$) that end in a call to $m_{or2}$. In this way, the analysis could not reveal that $m_{or2}$ is actually invoked by $m_{or1}$ [30].

- *Reflection* This technique uses the reflection capability of the Java language to replace direct method invocations with Java reflection methods that use class and method identifiers as parameters to perform the call. This makes actual method invocations difficult to inspect [30]. Listings 1 and 2 show an example of this transformation. In Listing 1, the method `m1` (of the class `MyObject`) is accessed through the operator "`.`" from the object instance, whereas in Listing 2 shows the same invoked method using the Java reflection API. In this example, a `java.lang.reflect.Method.invoke()` object is created (lines 2-3) and invoked (line 4) for a specific object instance (i.e., `obj`), whereas the class and method names are passed as parameters of these functions.

3.2.0.3 Encryption: This technique prevents accessing to parts or the entire code or resources (e.g., strings and asset files) of the app by using symmetric encryption algorithms. It involves storing the original code or resources in an encrypted form so that a decryption routine, inserted in the code, is invoked whenever an encrypted part needs to be accessed. The decryption key is stored somewhere in the APK or calculated at runtime. This technique introduces extra latency during app execution and severely complicates the analysis of the functionality of the encrypted part [27].

It is worth emphasizing that different obfuscation techniques can be combined to improve their effectiveness. For example, encrypting the strings of reflective calls can hide the method and class names invoked at runtime. This makes

Listing (1) Java direct method invocation

```
1    public class com.example.MyObject implements MyInterface {
2        public Object m1(Object prm) {}
3    }
4
5    public Object standardInvoke(com.example.MyInterface obj, Object arg) {
6        return obj.m1(param);
7    }
```

Listing (2) Java reflective method invocation

```
1    public Object refectionInvoke(com.example.MyInterface obj, Object arg) {
2        Class<?> cls = Class.forName("com.example.MyObject");
3        Method m = cls.getDeclaredMethod("m1");
4        return m.invoke(obj, param);
5    }
```

Listing (3) Java reflective method invocation with encrypted values

```
1    public Object refEncrInvoke(com.example.MyInterface obj, Object arg) {
2        String className  = decrypt("AXubduuiao...ZXW");
3        String methodName = decrypt("uibdadBUID...ncu");
4        Class<?> cls = Class.forName(className);
5        Method m = cls.getDeclaredMethod(methodName);
6        return m.invoke(obj, param);
7    }
```

it difficult to recover these values by static analysis of the apps. Listing 3 shows an example of the application of both obfuscation techniques to the code in Listing 1. In particular, the class and method names are decrypted at runtime (lines 2-3), hiding which methods are actually invoked. Note how these values are exposed only in an encrypted form, and could change if a different encryption key or algorithm was employed.

## 4 DATASET

For our experiments, firstly, we constructed a dataset with obfuscated and non-obfuscated apps. From this collection of apps and by means of static analysis, we obtain a set feature vectors that constitute the object of this study. This section describes how the app dataset is built and the types of features derived from the apps.

### 4.1 App Dataset

We build our app dataset using a subset of APKs from the AndroZoo repository [31], which contains more than 20 million of APKs with associated meta-data. This meta-data includes the source of the APK, the date, and the number of positive detections (VTD) in VirusTotal. Our objective was to obtain a dataset with the same number of malware and goodware samples, all of them free of obfuscation. We downloaded thousands of samples and filtered out those marked by APKiD [6] as "suspicious" of including obfuscation. To label samples we relied on the VTD values [32]: an app with VTD≥7 is considered malware, while an app with VTD=0 was considered goodware (apps with intermediate VTD values were filtered out).

In a second step, we generated obfuscated versions of the apps in the filtered dataset. To perform this process, we used the DroidChameleon [30], AAMO[33], and ObfuscAPK[29] tools. We chose these tools because (1) they are open source, (2) they provide a wide range of obfuscation techniques, and (3) they have previously shown to effectively evade Android malware detectors. Specifically, for each obfuscation tool, we try to obfuscate every app in the filtered dataset using six obfuscation techniques: Renaming, Junk Code Insertion, Reflection, Call Indirection and Encryption. The configuration of the tools was left as default for all techniques. The results of this process are summarized in Table 1.

Note that some tool combinations failed due to errors during the APK decompilation process. It is worth noticing that there were more failures in the case of malware apps than in goodware apps. ObfuscAPK was the tool with the best success rate, correctly obfuscating an average of 85% of the apps. On the contrary, we were unable to obtain obfuscated samples when trying to apply Encryption with AAMO, due to bugs introduced in the code by this tool that prevent the APK from being rebuilt. The attempts to use Renaming with DroidChameleon were also unsuccessful due to an error in the implementation of the tool. For other techniques, DroidChameleon and AAMO had average success rates of 55% and 28%, respectively. During this process, we realized that for some apps all the tool-technique combinations failed, and thus these apps were removed from the filtered dataset. As a result of this process, we obtained a "Clean" dataset which consists of 4 749 goodware and 4 067 malware (presumably) non obfuscated samples.

Table 2 summarizes the different datasets that will be used in the experiments. The criteria for the composition of these datasets will be explained in Section 5.

- NonObf: It includes the non obfuscated versions of the apps for which we could not obtain an obfuscated version with all the tools for at least one technique, i.e., apps that can be obfuscated using a specific tool and technique but not with the remaining tools using the same technique.
- CleanSuccObf: includes the subset of non obfuscated apps present in Clean, but not in NonObf. That is, all the apps for which all the tools have worked for at least one technique.
- The remainder datasets (Renaming, JCI, CallIndirection, Reflection, and Encryiption) contain the obfuscated versions of the apps in CleanSuccObf for that particular technique using all the tools.

---

6. https://github.com/rednaga/APKiD

TABLE 1: Success rate of different technique-tool obfuscation combinations for the apps in the Clean dataset. The first part of the name refers to the tool used to obfuscate the apps, with *DC* for DroidChamaleon, *AA* for AAMO, and *OA* for ObfuscAPK. The characters after the underscore refer to the strategy followed to obfuscate the apps: renaming (*rnm*), junk code insertion (*jcins*), call indirection (*ci*), reflection (*refl*) and encryption (*encr*).

| tool-technique | #Goodware samples | #Malware samples | Obf. Success Rate |
|---|---|---|---|
| DC_rnm | - | - | 0% |
| AA_rnm | 2244 | 1953 | 34% |
| OA_rnm | 5690 | 4317 | 81% |
| DC_jcins | 1855 | 1123 | 24% |
| AA_jcins | 2289 | 2019 | 35% |
| OA_jcins | 6003 | 4755 | 87% |
| DC_ci | 3664 | 2209 | 47% |
| AA_ci | 1337 | 1362 | 22% |
| OA_ci | 6050 | 4765 | 87% |
| DC_refl | 6200 | 3993 | 82% |
| AA_refl | 1332 | 1402 | 22% |
| OA_refl | 6080 | 4802 | 88% |
| DC_encr | 5008 | 3746 | 70% |
| AA_encr | - | - | 0% |
| OA_encr | 6074 | 4814 | 88% |

TABLE 2: Composition of datasets used in this work. The columns indicate the number of samples that comprise each set. The CleanSuccObf dataset contains the clean (original) apps for which we obtained obfuscated versions with all tools for at least one technique.

| Dataset | #Goodware samples | #Malware samples |
|---|---|---|
| Clean | 4749 | 4067 |
| NonObf | 1345 | 1211 |
| CleanSuccObf | 3404 | 2856 |
| Renaming | 3238 | 2868 |
| JCI | 1515 | 1008 |
| CallIndirection | 2118 | 1737 |
| Reflection | 2667 | 2484 |
| Encryption | 4790 | 4060 |

## 4.2 Feature Dataset

An app dataset has to be transformed into a dataset of feature vectors prior to perform malware detection using ML. Following a detailed literature analysis, we identified seven families of static analysis features that have proven to be useful for ML-based malware detection [6]. We used two well-known and widely used static analysis frameworks for Android to extract these features: Androguard

[34] and FlowDroid [35]. Sources of these features include: the *classes.dex* and *AndroidManifest.xml* files, as well as the contents of the *res* and *assets* directories of APKs.

### 4.2.1 Permissions

Permissions have commonly been used as a source of information for malware detection in Android [36–39]. In this category, we consider as features the full set of permissions provided by Google in the Android documentation [7], as well as the set of custom [8] permissions that developers may declare to enforce some functionality in their apps. Following this procedure, we extracted a set of binary features, each corresponding to the presence or absence of a given permission.

### 4.2.2 Components

An app consists of different software components that must be declared in the *AndroidManifest.xml* file. These elements have been widely used as a source of information for malware detectors [36, 38, 40, 41]. We extract a list of hardware and software components that can be declared using the ¡uses-feature¿ tag from the Android documentation [9], as well as every identifier for Activity, Service, ContentProvider, BroadcastReceivers and Intent Filters. In total, we obtained a set of 85 476 binary features, whose value is set to True or False for an app according to the presence of the feature in its *AndroidManifest.xml* file. We additionally derive seven frequency features accounting for the number of elements of each type in the app.

### 4.2.3 API functions

API libraries allow developers to easily incorporate additional functionality and features into their apps, being the main mean of communication between the programming layer and the underlying hardware. As such, analyzing the calls to methods of these libraries (API functions) constitutes a good instrument to characterize the functionality of apps, and, therefore, for malware detection. Following similar approaches to those proposed in the literature [36, 39, 40, 42], we extract a binary feature for each API method, and set its value to True if the app contains any call to that method within its code. In total, this set consist of 66 118 binary features.

### 4.2.4 Opcodes

The compiled Android code (Dalvik) consists of a sequence of opcodes. Opcode-based features provide insights about the code habits of developers as they represent fine-grained information about the functionality of apps [43]. Subsequences of opcodes, or simply $n$-grams, have been used for Android malware detection in [44–47]. Concerning the size of the subsequences, Jerome et. al [44] and Canfora et. al [45] observed that $n = 2$ offers a good trade-off between the size of the feature vector generated and the performance

7. https://developer.android.com/reference/android/Manifest.permission
8. https://developer.android.com/guide/topics/permissions/defining
9. https://developer.android.com/guide/topics/manifest/uses-feature-element.html

obtained by detectors. Therefore, we extract unique opcode subsequences of length 2 (or bi-grams) from the code of the apps, and create a feature to represent the number of appearances of each bigram in the code. The resulting vector contains a total of 25 354 frequency features.

### 4.2.5 Strings

The APK file strings are a valuable source of information for malware detection. In this regard, the most common strings include IP addresses, host names and URLs [36, 48]; command names [49, 50] and numbers [48]. We processed app files and found 2 425 892 unique strings. Following the procedure in [12], we observed that 98.5% of the strings were present in less than 1% of the samples. After removing these rare strings, we obtained 39 793 binary features, each representing the presence or absence of a specific string within the app files.

### 4.2.6 File related features

This type of features includes the size of code files and different file types inside the APK [16, 48, 49, 51]. We base our file type extractor on both, the extension of the file and the identification of the first bytes of the content (i.e., magic numbers) of files. The result is a new frequency feature for every unique combination of the extension (*ext*) and magic type (*mtype*), identified as *ext_mtype*. For files without extension, we use the complete file name instead. In total, this set consist of 65 986 frequency features per app.

### 4.2.7 Ad-hoc Features

As explained earlier, some specific detectors claim to use obfuscation-resistant features. We call the features used by these detectors that do not fall into any of the above categories ad-hoc features. They include: semantic features based on sink and source relationships in the code [50]; certificate information [16]; flags about the use of cryptographic, reflective, and command execution classes [42, 48, 51]; and resolved function names for native and reflective calls [18]. Due to the computational cost of obtaining these features, we limited the time spent computing them to 15 minutes per sample. The result is a set of 35 387 frequency features, each representing the number of occurrences of the feature within the app.

## 5 FEATURE VALIDITY

As a first step in this study, we have designed a set of experiments to determine the robustness and detection ability when obfuscation is present of the seven feature families described in the previous section. The first experiment analyzes the impact that different obfuscation strategies and tools have on the features. In the second experiment we evaluate the performance and stability of ML algorithms when using these features for malware detection.

### 5.1 Feature persistence

In this experiment, we aim to examine the impact of obfuscation on the features presented above. We analyze how and how much the features change in the presence of obfuscation. We highlight the disparities among obfuscation tools and how different implementation strategies to achieve the same obfuscation objective can affect the features.

To analyze these aspects, we calculate the feature *persistence* for each tool-technique obfuscation combination. This is done by determining the average level of overlap between the features of an original (clean) app and its obfuscated counterparts. To compute the feature overlap, we compare each pair of feature vectors calculated for an original app and its obfuscated version, and quantify the proportion of features with exact value matches. Note that for binary-featured representations (Permissions, Components, Strings and API functions), this is equivalent to computing the Jaccard index that measures the ratio between the shared elements and the total number of elements in the union of both feature vectors. Note also that, for frequency vectors, an increment or decrease in one unit or ten units has the same effect in this metric.

The results of this experiment are shown in Table 3. We find various degrees of persistence, in most cases over 0.8, with many exact matches between the feature vectors of clean and obfuscated APKs. Components and Permission features suffer the smallest changes when applying strategies such as Junk Code Insertion, Call Indirection, Reflection and Encryption (independently of the tool). Despite being affected by all techniques, File-Related features are also among the least affected on average. On the contrary, Ad-hoc, API functions and Opcode feature vectors change the most when obfuscation is applied. Nonetheless, the average persistence values for these features indicate that most fields (about 75%) are not affected by obfuscation. Therefore, in most cases, we conclude that the use of obfuscation is not reflected as a radical change in the feature vectors.

Persistence values refer to the proportion of features that remain unchanged, but do not tell us which particular features change the most when a tool-technique combination is applied. To shed some light on this regard, we selected the 15 features that change the most when obfuscation is applied. They may belong to different families. To obtain them, we measured the degree of discrepancy in the number of occurrences of each of these features, comparing the original application and the obfuscated version. To simplify the visualization, we show the results for each technique, averaging the discrepancy values for the three tools. The resulting rankings are shown in Figure 4. The name of each bar is the feature name (which includes its family). The number at the right of each bar is the degree of discrepancy, i.e., the average difference in the frequency of the feature between original and obfuscated versions of apps. Note that for easier interpretation, the scales are specific to each figure.

Regarding the persistence of the different feature families, Renaming mainly affected Components and API functions features, due to changes in the names of user-defined packages, classes, methods and fields. It also alters the declaration of custom permissions present in the code, since they depend on the name of the class where they are declared. However, as can be seen in Figure 4a, none of these features are among the 15 most affected, mainly because the names assigned to the classes are app-specific. In contrast, Opcode features are among those most significantly affected, due to changes in the order of methods when processing class files. This mainly changes the frequency of sequences that present

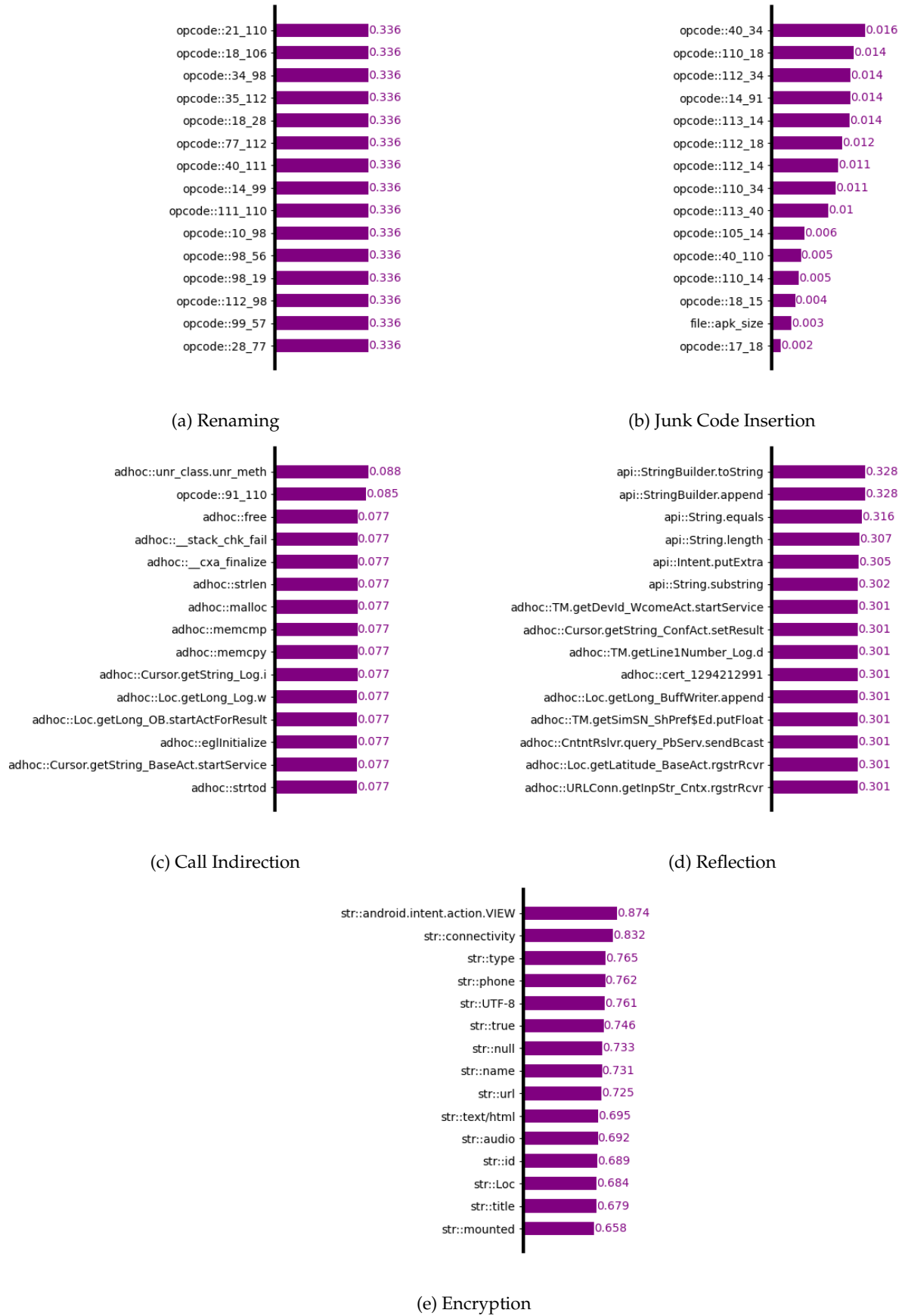

(a) Renaming

(b) Junk Code Insertion

(c) Call Indirection

(d) Reflection

(e) Encryption

Fig. 4: Top 15 of most changed features for each obfuscation strategy. The values on the right indicate the disparity or difference in the frequency of features between the obfuscated apps and their original versions. Results from the three tools have been averaged.

TABLE 3: Persistence of static analysis features when comparing clean and obfuscated apps using ObfuscAPK (OA), DroidChameleon (DC) and AAMO (AA).

| | Renaming | | | Junk Code Insertion | | | Call Indirection | | | Reflection | | | Encryption | | | Avg. |
|---|---|---|---|---|---|---|---|---|---|---|---|---|---|---|---|---|
| | OA | DC | AA | OA | DC | AA | OA | DC | AA | OA | DC | AA | OA | DC | AA | |
| **Permissions** | 0.972 | - | 1.0 | 1.0 | 1.0 | 1.0 | 1.0 | 1.0 | 1.0 | 0.932 | 0.799 | 1.0 | 1.0 | 1.0 | - | 0.977 |
| **Components** | 0.219 | - | 0.999 | 1.0 | 1.0 | 1.0 | 1.0 | 1.0 | 1.0 | 1.0 | 1.0 | 1.0 | 1.0 | 1.0 | - | 0.939 |
| **API functions** | 0.480 | - | 0.717 | 1.0 | 1.0 | 1.0 | 0.493 | 0.718 | 0.489 | 0.985 | 0.407 | 0.487 | 0.994 | 0.999 | - | 0.751 |
| **Opcodes** | 0.985 | - | 0.993 | 0.269 | 0.107 | 0.116 | 0.832 | 0.832 | 0.970 | 0.979 | 0.919 | 0.826 | 0.959 | 0.982 | - | 0.751 |
| **Strings** | 0.995 | - | 1.0 | 1.0 | 1.0 | 1.0 | 1.0 | 1.0 | 1.0 | 0.906 | 1.0 | 0.897 | 0.028 | 0.009 | - | 0.833 |
| **File Related** | 0.895 | - | 0.993 | 0.803 | 0.880 | 0.987 | 0.791 | 0.874 | 0.983 | 0.845 | 0.904 | 0.987 | 0.926 | 0.923 | - | 0.907 |
| **Ad-hoc** | 0.923 | - | 0.942 | 0.655 | 0.345 | 0.560 | 0.455 | 0.667 | 0.409 | 0.888 | 0.859 | 0.850 | 0.92 | 0.922 | - | 0.722 |

invocation instructions (opcodes 110, 111 and 112).

In concordance with persistence values in Table 3, Figures 4c and 4b show that Call Indirection and Junk Code Insertion techniques are particularly detrimental for features based on code information, with Opcode and Ad-hoc features being the most sensitive to both types of obfuscation. In particular, Ad-hoc features are the most affected by Call Indirection (see Figure 4c) due to the added complexity in the analyses required for their extraction. This is the case of sink and source relations between API functions such as *Cursor.getString* and *Log.i*. Also, due to the addition of indirect calls, this technique increases the frequency of some opcode sequences such as "90 110" formed by an *iput* (90) instruction followed by an invoke*invoke* (110). This technique involves adding hundreds of auxiliary (indirect caller) methods per class, either in separate or in the API classes inside the API. However, these methods are randomly named, which limits their impact (their popularity will be low). As shown in Figure 4b, Junk Code Insertion greatly alters the frequency of most Opcode sequences due to the inclusion of useless instructions, mainly *goto* (40) and *invoke* (110, 112, 113). The introduction of useless code also greatly impacts on the size of the APK file (File-related feature *file:apk_size*).

Reflection changes the persistence of features extracted from code analysis. This effect is clearly perceptible in Figure 4d. With this technique, the code is modified to hide the originally called methods and use reflective calls instead. This mainly affects the API functions that are called more frequently in the code, including string-related functions such as *toString*, *append*, *equals*, or *length*. Ad-hoc functions are among the most changed due to the added complexity of identifying sink and source relationships that contain reflective code. Reflection also results in new string features that contain the class and function names invoked by reflection. However, these are declared once in the code, so their frequency is kept low. Permission features are affected because Reflection can hide the presence of protected API functions that require specific permissions to be granted.

Encryption adds helper classes with the decryption routines that are used to hide user-defined strings and parameters. Therefore, API, Opcode, Ad-hoc, and File-related features are affected by the modifications introduced in the code. However, the main target of this technique are String features, as illustrated in Figure 4e and Table 3. These are heavily affected because their original values are encrypted. In this regard, the top 15 features most changed by encryption are strings related to the app's user interface (`UTF-8`,`phone`, `id`, `title`, `type`).

### 5.1.1 Differences between Obfuscation Tools

As seen in Table 3, changes in features depend on the tool used. These differences are due to implementation particularities. Indeed, because of these peculiarities, obfuscation can even alter features that are not primary target of the chosen obfuscation technique. Figure 5 depicts the average level of overlap between the features obtained for the same apps when obfuscated using different tools. Darker colors indicate less overlap in the obtained feature vectors, while lighter colors represent higher agreement. To better explain the differences obtained, we manually examined the code of these obfuscation tools as well as the features obtained from different samples. Due to space limitations and in order to make this paper more readable, we omit very specific implementation details and limit our discussion to the more prominent differences.

We observed that of all the tools analyzed, none of them considered parameter randomization when implementing the different obfuscation techniques except for Junk Code Insertion. As such, for a given tool, the extracted feature vectors are only dependent on the input (app data). In other words, given the same input, a particular tool-technique combination will always return the same output. It is worth mentioning that all tools obfuscate (modify) the Android or Java libraries when this type of content is included in the APK, mainly due to poor checks during obfuscation. Since this code is not user-related, such changes may break the execution flow of apps.

The largest differences between obfuscation tools are present for API function and Ad-hoc features. This aspect is clearly perceptible in Figures 5. In the case of Reflection, we noticed that ObfuscAPK and AAMO perform a fine-grained checking when selecting the set of candidate function calls to be transformed, so that errors introduced by obfuscation are minimal. In contrast, DroidChameleon obfuscates calls whose package matches any of the prefixes included in a pre-defined list, without making any additional checks. In consequence, as shown in Figures 5a and 5c feature overlap is low since DroidChameleon results in a higher number of transformed API calls with respect to ObfuscAPK and AAMO.

The way in which the files to be transformed are selected is also the explanation behind the differences observed between AAMO and ObfuscAPK with Renaming (see Figure 5b). By default, ObfuscAPK selects all the files within the APK as candidates for Renaming. This translates into changes in the content of files even if they belong to the Java library or the Android framework. Hence, features in the Components and API functions families are greatly modified by this tool. In contrast, AAMO performs some additional checks aiming to avoid modifying this type of content and presents a reduced impact on these features. Nonetheless, as shown by the persistence values for API-related features in Table 3, these checks are insufficient. For example, classes that are part of the "com.android" package are obfuscated because they do not match the name of the AAMO blacklisted "android" package.

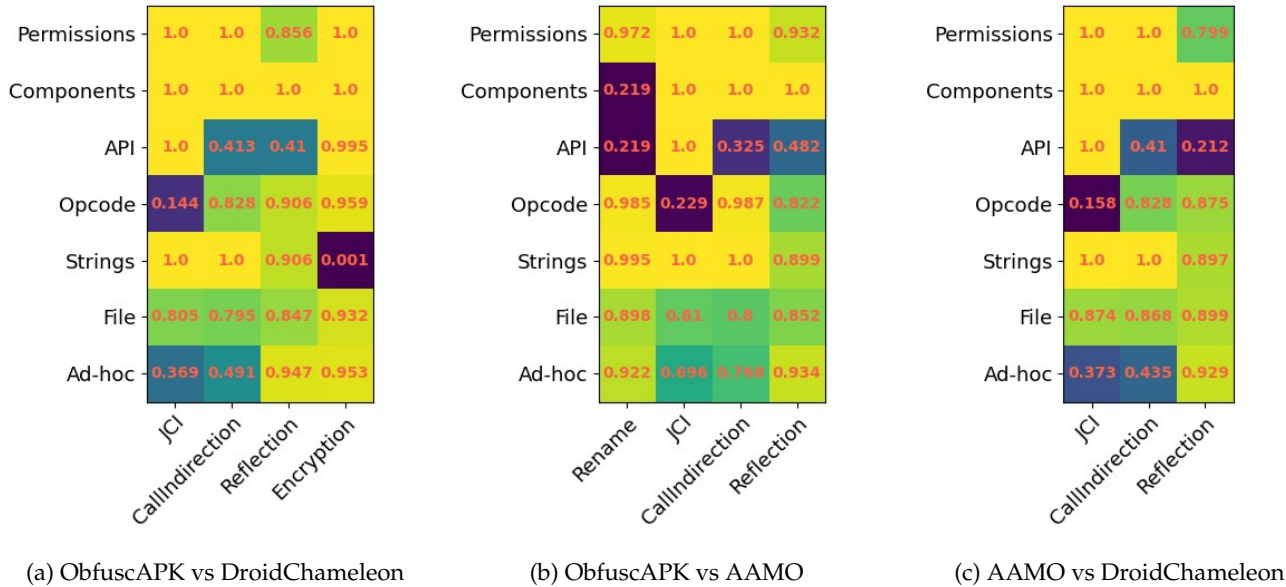

(a) ObfuscAPK vs DroidChameleon

(b) ObfuscAPK vs AAMO

(c) AAMO vs DroidChameleon

Fig. 5: Feature overlap between every pair of obfuscation tools using different obfuscation strategies. Note that because Rename and Encryption do not work for DroidChameleon and AAMO, respectively, the corresponding columns are omitted.

The disparities observed for API function features with CallIndirection between the three tools (see Figures 5a, 5b and 5c) are due to the way intermediate methods are created. ObfuscAPK and AAMO insert the code of intermediate methods within the class file of the original calling method, whereas DroidChameleon adds this code to a (separate) helper class. As a result, methods inserted by ObfuscAPK and AAMO inside the class files of the Android framework are considered as API features by the feature extraction process. Since these tools use different naming conventions for these new methods, the resulting features do not overlap.

When applying encryption, ObfuscAPK and DroidChameleon use different algorithms and parameters. This explains the differences observed between both tools in Figure 5a. In particular, ObfuscAPK uses the AES cipher for encryption, whereas DroidChameleon uses Caesar's algorithm. In both cases, the encryption key is hardcoded in their respective code.

## 5.2 ML Performance

We devise a second set of experiments to 1) analyze the ability of static features to detect malware, and 2) study the stability of these features for malware detection within ML algorithms in the presence of obfuscation. In all experiments, the RandomForests classification algorithm is used without any parameter optimization [52], as implemented in scikit-learn, a widely-used python library for ML [53].

The first scenario is focused on analyzing the predictive power of the different feature families in a fully non-obfuscated (clean) environment. For model training we use the NonObf dataset[10]. For evaluation purposes we used the apps from the CleanSuccObf dataset. Our objective is to evaluate the ability of an off-the-shelf classifier to approximate the class $y$ of apps (malware or goodware) as a function of the original (non-obfuscated) features $x_{orig}$ obtained from clean apps.

Table 4 shows the performance of the trained models. As can be seen, most features present high true positive rates (TPR above 0.8) and moderately low false positive rates (FPR below 0.2). Therefore, it can be concluded that most feature families provide enough information to enable effective malware detection using ML algorithms. This is particularly true for API functions and String features. On the contrary, the model generated using File-Related features performs similar to a random choice model (an $A_{mean}$ value of 0.5) and therefore, we can say that these features are not suitable for the purpose at hand.

Even with high persistence values, small changes in feature vectors can lead to large changes in the performance of an ML algorithm. This may be the case if the small set of changed features is the most informative for a classifier and strongly influences its prediction. Consequently, this second scenario investigates the sensitivity of ML algorithms to the changes induced by feature vector obfuscation.

We use the ML models trained in the previous experiment (i.e., with clean apps from the NonObf set) and compile two separate evaluation sets for each obfuscation strategy. The first set, known as the obfuscated evaluation set, consists of (obfuscated) samples from the corresponding Renaming, JCI, CallIndirection, Reflection or Encryption datasets. The other set comprises the clean versions of those apps in the obfuscated dataset. By comparing the predictions made by the ML model for the clean and obfuscated versions of the same app, we can assess whether or not obfuscating an app can change the decision made by the ML model. We leverage the Jaccard index to compute the overlap between the predictions for the clean and obfuscated apps, and we refer to this measure as *insensitivity*. Thus, a high level of insensitivity indicates that the predictions made by a model are preserved even when obfuscation is applied to the apps.

---

10. Note that an error during the obfuscation process of an app from this set for a given tool can be due to an error in the obfuscation tool, since the same app has been successfully obfuscated using other tools for the same and other strategies.

TABLE 4: Performance of static analysis features for malware detection using non-obfuscated apps for both training and evaluation. TPR stands for the True Positive Rate, i.e., the number of malware correctly identified. FPR stands for the False Positive Rate, i.e., the number of goodware erroneously identified as malware. The $A_{mean}$ is the average of the TPR and the True Negative Ratio (1-FPR).

| | TPR | FPR | $A_{mean}$ |
|---|---|---|---|
| **Permissions** | 0.867 | 0.156 | 0.855 |
| **Components** | 0.808 | 0.157 | 0.825 |
| **API functions** | 0.928 | 0.081 | 0.923 |
| **Opcodes** | 0.884 | 0.252 | 0.816 |
| **Strings** | 0.907 | 0.082 | 0.912 |
| **File Related** | 0.265 | 0.197 | 0.534 |
| **Ad-hoc** | 0.768 | 0.143 | 0.812 |

The measured insensitivity values are compiled in Table 5. As can be seen, the decisions of ML models for most feature families are consistent. Permissions, Components or API functions result in stable predictions regardless of the obfuscation status of the apps, with insensitivity levels exceeding 90%. On the contrary, Ad-Hoc, Opcode and File-Related features exhibit high fluctuations in the decisions made by the models. This suggests a greater sensitivity of models to changes introduced by obfuscation in these features.

We wondered if the persistence of features when obfuscation is applied is somehow related to the insensitivity of ML models based on those features. In Figure 6, we represented the persistence and insensitivity values for the different feature families. As can be seen, in general, there is a high correlation between low persistence and high sensitivity, meaning that larger changes in the features vectors induce larger changes in the predictions of the ML algorithm. See for example Opcode features with JCI in Figure 6b and String features with Encryption in Figure 6e). However, high persistence values do not necessarily mean that the retained features are the ones that are more helpful to the ML models in making accurate predictions. For example, Ad-hoc features show high persistence values when applying Reflection (changed features are only 16% of the total). Still, the insensitivity is rather low, indicating those include the features that play an important role in the accuracy of predictions (see Figure 6d). Another example is File-related features, which show the most irregular behavior for ML models despite the small proportion of features altered by obfuscation (10% on average as shown in Figure 6f). In this regard, in the previous scenario we evidenced that File-related features lack informativeness for detection (see Table 4).

The previous results highlight an important finding: while persistent features are commonly considered reliable predictors for malware detection, persistency is not the sole factor influencing the robustness of the detection model. On the contrary, high insensitivity values implicate a high persistence on features, so it is a more adequate indicator of robustness. Therefore, it is crucial to carefully examine the impact of obfuscation-induced changes on the informativeness of the features, as even small changes can significantly impact prediction performance. In the next section, we explore the selection of different feature vectors based on ML performance and feature insensitivity to changes to develop robust malware detection models.

## 6 ROBUST MALWARE DETECTION

We hypothesize that it is possible to build a robust classifier (one with accurate predictions when dealing with both clean and obfuscated apps) by using features that are both relevant (generate good models with clean apps) and insensitive (the decision of the classifier does not change between the clean and the obfuscated version of an app). We call these *robust* features. On the contrary, features that obtained low insensitivity values (i.e., are highly sensitive) or are irrelevant for ML models are prone to cause fluctuations in the predictions of ML models when obfuscation is used.

To select a set of robust features based on the previous statements, we use the $A_{mean}$ metric reported in Table 4 and the average feature insensitivity reported in Table 5. Specifically, we set three thresholds (0.8, 0.85, and 0.9) for both metrics ($A_{mean}$ and feature insensitivity values) as criteria for selecting different sets of robust features – to be used for training and testing ML-based detectors. Table 6 shows the feature families selected at each threshold: the strictest threshold selects only the API functions family (A), the intermediate threshold selects API functions, Permissions, and Strings (PAS), and the lower threshold selects Permissions, API functions, Components, and Strings feature families (PACS). Given the large number of features in the selected groups (particularly in PAS and PACS), we rank the features of each family based on the relevancy value computed by the corresponding RandomForest models from Section 5.2, and select only the best 2 000 features of each family. This selection does not apply to the Permissions family, as it only includes 683 features.

Three RandomForest classifiers are trained using the apps in the NonObf dataset (which does not include obfuscated samples): one for A features, another one for PAS features and the third one for PACS features. For evaluation, we use the CleanSuccObf dataset for the non-obfuscated scenario, whereas for the obfuscated scenario, we use the apps from the Renaming, JCI, CallIndirection, Reflection and Encryption datasets. The prediction performances of these three models are summarized in Table 7. As expected, results are good for the tests without obfuscation, with true positive rates over 90% and low ratios of false positives (under 8%). When tested with obfuscated apps, the model trained with API function features (A) showed a performance reduction of 3% in terms of $A_{mean}$ with respect to the use of the non-obfuscated versions of the same apps, mainly due to the effect of Reflection. The use of additional features (PAS, PACS) seems to provide valuable information to models: with PAS features, we can observe a 2% reduction in the number of false positives, while the ability to correctly detect malware increases 3% with respect to the performance numbers obtained for the model using API functions. The addition of Components features (PACS) did not improve the performance with respect to PAS, obtaining 2% fewer true positives and a reduction of 1% in the number of obfuscated goodware being misclassified. Therefore, the best model is PAS, the one that uses Permissions, API functions and Strings.

TABLE 5: Feature insensitivity, i.e., the overlap between the classifications made by the ML models for original and their obfuscated variants using ObfuscAPK (OA), DroidChameleon (DC) and AAMO (AA).

| | Renaming | | | Junk Code Insertion | | | Call Indirection | | | Reflection | | | Encryption | | | Avg. Over. |
|---|---|---|---|---|---|---|---|---|---|---|---|---|---|---|---|---|
| | OA | DC | AA | OA | DC | AA | OA | DC | AA | OA | DC | AA | OA | DC | AA | |
| Permissions | 0.986 | - | 1.0 | 1.0 | 1.0 | 1.0 | 1.0 | 1.0 | 1.0 | 0.934 | 0.674 | 1.0 | 1.0 | 1.0 | - | 0.968 |
| Components | 0.506 | - | 1.0 | 1.0 | 1.0 | 1.0 | 1.0 | 1.0 | 1.0 | 1.0 | 1.0 | 1.0 | 1.0 | 1.0 | - | 0.961 |
| API functions | 0.986 | - | 0.992 | 1.0 | 1.0 | 1.0 | 1.0 | 1.0 | 1.0 | 0.946 | 0.305 | 0.998 | 1.0 | 1.0 | - | 0.939 |
| Opcodes | 0.950 | - | 0.964 | 0.446 | 0.235 | 0.087 | 0.899 | 0.963 | 0.900 | 0.922 | 0.954 | 0.893 | 0.923 | 0.940 | - | 0.774 |
| Strings | 1.0 | - | 1.0 | 1.0 | 1.0 | 1.0 | 1.0 | 1.0 | 1.0 | 1.0 | 1.0 | 1.0 | 0.296 | 0.078 | - | 0.874 |
| File Related | 0.027 | - | 0.805 | 0.013 | 0.052 | 0.384 | 0.013 | 0.072 | 0.380 | 0.025 | 0.106 | 0.575 | 0.091 | 0.030 | - | 0.197 |
| Ad-hoc | 0.624 | - | 0.813 | 0.706 | 0.012 | 0.583 | 0.434 | 0.751 | 0.415 | 0.621 | 0.370 | 0.238 | 0.720 | 0.741 | - | 0.540 |

TABLE 6: Features selected for robust malware detection based on different thresholds for the $A_{mean}$ and feature insensitivity.

| Threshold | Feature types | #Features |
|---|---|---|
| 0.8 | Permissions, API functions, Components, Strings | 6 683 |
| 0.85 | Permissions, API functions, Strings | 4 683 |
| 0.9 | API functions | 2 000 |

TABLE 7: Performance of different robust feature combinations for ML malware detection. A, stands for the model using exclusively API functions. PAS, refers to proposal using Permissions, API functions and Strings, whereas PACS uses Permissions, API functions Components and Strings.

| | Non-Obfuscated | | | Obfuscated | | |
|---|---|---|---|---|---|---|
| | A | PAS | PACS | A | PAS | PACS |
| TPR | 0.928 | 0.920 | 0.930 | 0.858 | 0.889 | 0.876 |
| FPR | 0.081 | 0.065 | 0.065 | 0.060 | 0.044 | 0.035 |
| $A_{mean}$ | 0.923 | 0.927 | 0.932 | 0.898 | 0.922 | 0.914 |

For comparison, we also evaluate the performance of our RandomForest classifier with PAS features against Reveal-Droid, a robust state-of-the-art malware detector [18], and Drebin, a high-performing detector [36]. Both detectors use their own sets of static analysis features and ML algorithms. RevealDroid features include API function and package counts, native calls extracted from binary executables and function names resolved from reflective and dynamic code loading calls. These account for a total of 59 072 features that are used to train a RandomForest model to perform malware detection. The features used by Drebin comprise declared and requested permissions, app components, hostnames, IPs, commands and suspicious and restricted API functions. This totals 253 881 binary features that are used to train a linear Support Vector Machine (SVM) goodware-malware classifier.

Table 8, shows the performance of our best model (RandomForest with PAS features) against RevealDroid and Drebin. As can be seen, PAS outperformed both state-of-the-art detectors for the non-obfuscated and obfuscated scenarios. With obfuscated apps, our robust proposal presented a 1% and a 6% higher detection rate, and 4% and 8% lower malware misclassification, with respect to Drebin and RevealDroid. These good numbers demonstrate that using a small set of Permissions, API functions and Strings is enough to perform malware detection in Android. In contrast to RevealDroid and Drebin, which mainly rely on Strings and API features, PAS considers a more balanced feature vector and hence, it is more robust against different obfuscation strategies. This experiment demonstrates that obfuscated malware and goodware can be identified using off-the-shelf ML algorithms and features obtained from static analysis, even without providing these algorithms with any information about the strategy or tool used to obfuscate apps.

# 7 DISCUSSION AND FUTURE WORK

The experiments carried out in this paper evidence that, as commonly assumed [54], static analysis features can be affected by specific obfuscation techniques. On one hand, feature persistence showed that all the feature families are affected by at least one obfuscation technique. Among them, the features obtained from the manifest of applications proved to be the most stable. Nonetheless, and contrary to what is commonly argued [55], our experiments also demonstrate that static analysis features can be a reliable source of information for ML malware detection. In this regard, we observed that some obfuscation strategies can result in additional features while leaving the original features unaltered. For example, this is the effect of CallIndirection in API functions, or Reflection in Strings. In most cases, the impact of obfuscation is limited to less than 20% of all the features derived from the samples (for example, at most 20% of the features are affected by Call Indirection and about 15% of them are altered by Reflection). An interesting line of research in this regard could be to analyze whether static analysis frameworks have flaws that magnify the risk of obfuscation. This aspect would help developers to improve static analysis tools and also facilitate practitioners to select the most reliable tool.

We also observed that the alterations caused by obfuscation on the features vary significantly between different tools, mainly due to implementation particularities. However, the lack of randomization in these tools makes them produce the same output for the same input value, even for different source apps or executions of the same tool. Such behavior is useful to hide the explicit information provided by, for example a class name, but it is insufficient to conceal the intrinsic information, i.e., relationships between features, such as correlations. This means that apps that contain a similar characteristic, when obfuscated using the same tool, will maintain a similar relations between the obfuscated values than between the original (unobfuscated) features. Additionally, most obfuscators need to improve the implementation of some obfuscation techniques due to the high failure rates they present. From the point of view of the users of these tools, these aspects pose a limitation of obfuscators. Therefore, the proposition and implementation of better obfuscation strategies and tools for Android is a promising research area. Such strategies should be accompanied with evaluations to ensure that the execution of the applications is not broken.

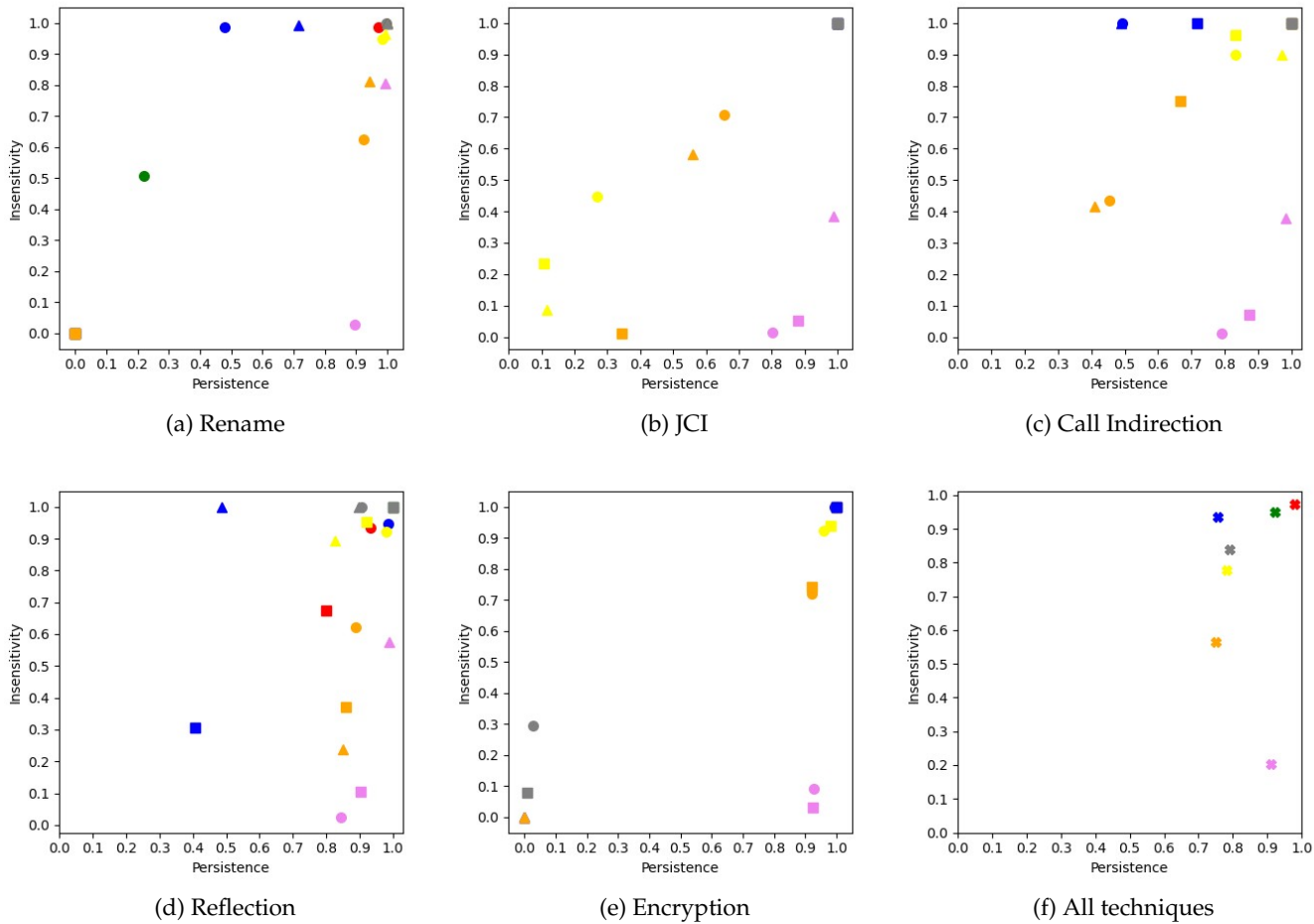

(a) Rename  (b) JCI  (c) Call Indirection

(d) Reflection  (e) Encryption  (f) All techniques

Fig. 6: Relation between persistence and insensitivity to changes of the different features for each obfuscation technique and tool. Every color makes reference to a feature family, with red for Permissions, green for Components, blue for API functions, yellow for Opcodes, gray for Strings, violet for File-Related, and orange for Ad-hoc features; whereas symbols make reference to the values reported on each feature family for ObfuscAPK (circles), AAMO (triangles) and DroidChameleon (squares). The average of all tools is represented by the croos symbol.

TABLE 8: Performance of robust ML detectors based on static analysis features. PAS refers to our robust detection proposal using Permissions, API functions and Strings.

| | Non-Obfuscated | | | Obfuscated | | |
|---|---|---|---|---|---|---|
| | RevealDroid | Drebin | PAS | RevealDroid | Drebin | PAS |
| **TPR** | 0.856 | 0.914 | 0.920 | 0.832 | 0.876 | 0.889 |
| **FPR** | 0.117 | 0.103 | 0.065 | 0.12 | 0.088 | 0.044 |
| **A$_{mean}$** | 0.869 | 0.905 | 0.927 | 0.856 | 0.893 | 0.922 |

The performance analysis of ML models trained with static analysis features revealed that some feature types (families) typically proposed for malware detection [56], such as file-related features, are not effective in differentiating malware. This experiment, conducted on non-obfuscated applications, identified API functions and Strings as the most informative features for malware detection, achieving detection rates of over 90% and a low false positive rates of 8%. We analysed the impact of obfuscation-induced changes on the informativeness of features and showed that even small changes can have a significant impact on performance. Therefore, feature persistence should not be considered as the sole criterion for robust malware detection. This finding demystifies a common assumption in the Android malware detection field, which is to consider highly persistent features as robust.

By combining features that exhibited high insensitivity to changes and presented high ML accuracy with non-obfuscated apps, we demonstrated that ML-based malware detection using static analysis features can be robust in the presence of obfuscation. Remarkably, this remains true even in scenarios where no knowledge about the obfuscation techniques applied to the apps is assumed, i.e., the obfuscated data is not taken into account during the training process. In this scenario, our proposed robust detection approach based on a stock classifier outperformed Reveal-Droid, the current state-of-the-art obfuscation-resilient detector, and Drebin, the best proposal for malware detection in Android according to a recent comparative [15]. Specifically, our detector achieved 92% of correct classifications, compared to 89% and 85% for Drebin and RevealDroid, respectively. This result shows that Android malware detectors go beyond the selection of a ML algorithm. Therefore, instead of focusing on the ML aspect, richer and more robust app representations that benefit from the integration of different static analysis data should be further explored.

As a final note, we are aware that some limitations apply to the work carried out for this paper. The main one is that our analysis is limited to individual obfuscation strategies. However, these strategies can be combined in order to increase the probability of circumventing detectors. Also, the order in which these obfuscation strategies are combined influences the results obtained (note that the information hidden by a previous obfuscation technique becomes invisible for the next obfuscation strategy). Evaluating all these combinations would drastically increase the number scenarios to be evaluated. The cost of the required experimentation would be unfeasible since: (1) samples would have to be obfuscated combining strategies and tools, and (2), feature extraction and model training would have to be performed for the resulting obfuscated samples. Moreover, in our opinion, such extensive analysis would also hinder to clearly conclude the impact caused by each individual obfuscation strategy. In this regard, this work can be seen as a first step in investigating the impact that the combination of different obfuscation strategies can have on static analysis features. As for future work, we plan to extend our experiments to additional obfuscation techniques, such as packing.

## 8 Conclusions

This paper delved into the effectiveness of static analysis features for ML-based Android malware detection in the presence of obfuscation. To perform this assessment, we generated a variety of datasets by applying different obfuscation strategies to apps with the help of three state-of-the-art obfuscators. Seven families of static analysis features were defined and evaluated throughout an extensive set of experiments. We identified which families are more persistent when obfuscation is applied, and which families are more informative for correct Android malware detection. Based on these findings, we proposed the use of Permissions, API functions and Strings for ML-based malware detection. A stock implementation of the RandomForest classification algorithm using these robust features was used to generate a ML model able to separate malware from goodware with a remarkable success rate, without any prior knowledge of the specific obfuscation techniques applied to apps. In particular, this detector correctly identified 89% of evasion attempts with a low false positive rate of 4%, outperforming the current state-of-the-art solution for obfuscation-resistant Android malware detection.

## Acknowledgements

This work has received support from the following programs: PID2019-104966GB-I00AEI (Spanish Ministry of Science and Innovation), IT-1504-22 (Basque Government), KK-2022/00106 (Elkartek project supported by the Basque Government). Borja Molina-Coronado holds a predoctoral grant (ref. PRE_2021_2_0230) by the Basque Government.

**Borja Molina-Coronado** received his B.Sc. in Computer Engineering and his M.Sc. in Computer Science from the Technical University of Valencia and the University of the Basque Country UPV/EHU, in 2015 and 2017, respectively. He is a Ph.D. student in the Dept. of Computer Architecture and Technology of the UPV/EHU. His main research interest include malware analysis, network security and machine learning.

**Antonio Ruggia** Antonio Ruggia is a Ph.D. student in Security, Risk, and Vulnerability at the University of Genoa since November 2020. He is interested in several security topics, including Mobile Security, with a specific interest in Android, malware, and data protection. He graduated in October 2020 from the University of Genoa and participated in the 2019 CyberChallenge.it, an Italian practical competition for students in Cybersecurity. Since 2018, he has worked as a full-stack developer in a multinational corporation.

**Usue Mori** received her M.Sc. Degree in Mathematics, and a Ph.D. in Computer Science from the University of the Basque Country UPV/EHU, Spain, in 2010 and 2015, respectively. Since 2019, she has been working as a lecturer in the Dept. of Computer Science and Artificial Intelligence of the University of the Basque Country UPV/EHU. Her main research interests include clustering and classification of time series.

**Alessio Merlo** Alessio Merlo (Senior Member, IEEE) received the Ph.D. degree in computer science from the University of Genoa in 2010. He is currently a Full Professor in computer engineering with the Centre for Higher Defence Studies (CASD), Rome, Italy. He has published more than 120 scientific papers in international conferences and journals. His research interests include mobile security, where he contributed to discovering several high-risk vulnerabilities both in applications and the android OS and system security.

**Alexander Mendiburu** is a full professor at the Dept. of Computer Architecture and Technology of the University of the Basque Country UPV/EHU, where he has been working since 1999. He received his B.Sc. Degree in Computer Science and his Ph.D. Degree from the University of the Basque Country, Spain, in 1995 and 2006, respectively. His main research areas are evolutionary computation, time series, probabilistic graphical models, and parallel computing.

**Jose Miguel-Alonso** is a full professor at the Dept. of Computer Architecture and Technology of the University of the Basque Country UPV/EHU. Formerly, he was a Visiting Assistant Professor at Purdue University. He received his M.Sc. in Computer Science in 1989 and his Ph.D. in Computer Science in 1996, both from the UPV/EHU. He carries out research related to parallel and distributed systems, in areas such as network security, software security, performance modeling and resource management in large-scale computing systems. Prof. Miguel-Alonso is a member of the IEEE Computer Society and of the HiPEAC European Network of Excellence.